# Single-chip photonic deep neural network for instantaneous image classification

**Farshid Ashtiani, Alexander J. Geers, and Firooz Aflatouni\***

*Department of Electrical and Systems Engineering, University of Pennsylvania, Philadelphia, PA 19104, USA*
*\*firooz@seas.upenn.edu*

**Deep neural networks with applications from computer vision and image processing to medical diagnosis[1-5] are commonly implemented using clock-based processors[6-14], where computation speed is generally limited by the clock frequency and the memory access time.**

**Advances in photonic integrated circuits have enabled research in photonic computation[15-17], where, despite offering excellent features such as fast linear computation, no integrated photonic deep network has been demonstrated to date due to the lack of scalable on-chip nonlinear functionality and the loss of photonic devices, making scalability to a large number of layers challenging.**

**Here we report the first integrated end-to-end photonic deep neural network (PDNN) that performs instantaneous image classification through direct processing of optical waves. Images are formed on the input grating coupler array (pixels). Optical waves impinging on different pixels are coupled into nanophotonic waveguides and processed as the light propagates through layers of neurons on-chip. Each neuron generates an optical output from input optical signals, where linear computation is performed optically and the nonlinear activation function is realised opto-electronically. The output of a laser coupled into the chip is uniformly distributed among all neurons within the network providing the same per-neuron supply light. Thus, all neurons have the same optical output range enabling scalability to a deep network with a large number of layers. The PDNN chip is used for 2- and 4-class classification of handwritten letters achieving accuracies of higher than 93.7% and 90.3%, respectively, with a computation time less than a single clock cycle of state-of-the-art digital computation platforms. Direct clock-less processing of optical data eliminates photo-detection, analogue-to-digital conversion, and the requirement for a large memory module, enabling significantly faster and more energy-efficient neural networks for the next generations of deep learning systems.**

Artificial deep neural networks are employed in a growing number of applications such as pattern recognition[1], natural language processing[2], and medical diagnosis[3-5]. Inspired by the distributed data processing in the human brain, such networks are designed to process the input data using interconnected layers of neurons (nodes) which can be trained using a set of training data to learn a specific task. Once trained, the network can be used to perform the same task on a new set of data with high accuracies. Figure 1a shows a general architecture of a deep neural network, where the input data is first arranged and then processed using the neurons of the first layer followed by the intermediate (hidden) layers. The classification result appears at the output of the last layer, the output layer. Each neuron within the network generates an output by passing the weighted-sum of its inputs through a nonlinear activation function (Fig. 1b).

Deep neural networks are usually implemented using digital clock-based platforms such as graphics processing units (GPUs)[13,14] or application specific integrated circuits (ASICs)[19,20]. GPUs are highly reconfigurable processors that are capable of performing a large number of computations in parallel, yet, their computation time is mainly limited by the clock frequency (mostly less than 3 GHz for the state-of-the-art GPUs) and the memory access time[21]. Implementation of deep networks using ASICs can provide one to two orders of magnitude improvement in terms of performance per unit energy consumption compared to GPUs[22], however, they generally face similar challenges as GPUs, which become more significant for more complex networks with a large number of neuron layers. Furthermore, for digital implementation platforms, the input data often needs to be converted to the electrical domain, digitized, and processed. Often a large memory unit is required to store the data set, which limits the processing time and, in the case of image or video classification, may present privacy implications.

Large bandwidth available at optical frequencies as well as low propagation loss of nanophotonic waveguides (serving as interconnects) make photonic integrated circuits a promising platform to implement fast and energy-efficient processing units[15-18] that can augment the performance of conventional digital processors. Recently, photonic implementations of deep neural networks have been reported[15-17, 23-30] that offer key features such as instantaneous linear operation and low-loss high bandwidth connectivity within the network. However, all demonstrations of neural networks to date have been limited to either bench-top setups[28-30] or integration of parts of a deep learning network[15-17,23-27] and due to the lack of scalable on-chip nonlinear functionality and uncompensated loss of cascaded photonic devices, no scalable fully integrated photonic deep learning system for data classification has been demonstrated.

Here we report the demonstration of the first integrated end-to-end photonic deep neural network that utilises computation-by-propagation to perform instantaneous image classification. Target images are formed on an array of grating couplers serving as input pixels of the PDNN chip. Optical waves impinging on different pixels are coupled into the corresponding nanophotonic waveguides and processed as the light propagates through neurons of different layers on the PDNN chip. All neurons within the network have the same optical output range enabling scalability to a deep network with a large number of layers. As a proof of concept, the PDNN chip was used for 2- and 4-class classification of handwritten letters achieving accuracies of higher than 93.7% and 90.3%, respectively, while the computation time is less than a single clock cycle of the state-of-the-art digital computation platforms. As a point of comparison, a conventional deep neural network classifier implemented in Python environment using Keras[31] achieves 96% accuracy for the same data set. The implemented PDNN features direct clock-less processing of input images which eliminates the need for photo-detection, scaling and amplification, analogue-to-digital conversion, data alignment, and a large memory module enabling the realization of significantly faster and more energy-efficient, yet privacy-aware neural networks for the next generations of deep learning systems. The PDNN chip was integrated within a footprint of 9.3 mm$^2$.

The architecture of the implemented photonic deep neural network chip and the corresponding N-input photonic neuron are illustrated in Figs. 1c and 1d, respectively. The target image is formed on the input 5x6 pixel array, which is divided into four overlapping 3x4-pixel sub-images (shown in dark blue, red, light blue, and light green in Fig. 1c). Input nanophotonic waveguides are arranged to route pixels of each sub-image to a 12-input neuron within the input layer forming a convolution layer[6,7]. Convolution layers are commonly used within a deep network in image/pattern recognition applications enabling lower number of connections and a more efficient feature extraction[8-10]. The outputs of the 1st layer are fully-connected to the three neurons of the 2nd layer. Similarly, the three outputs of the 2nd layer are fully-connected to the two neurons of the 3rd layer, generating two network outputs, Out1 and Out2.

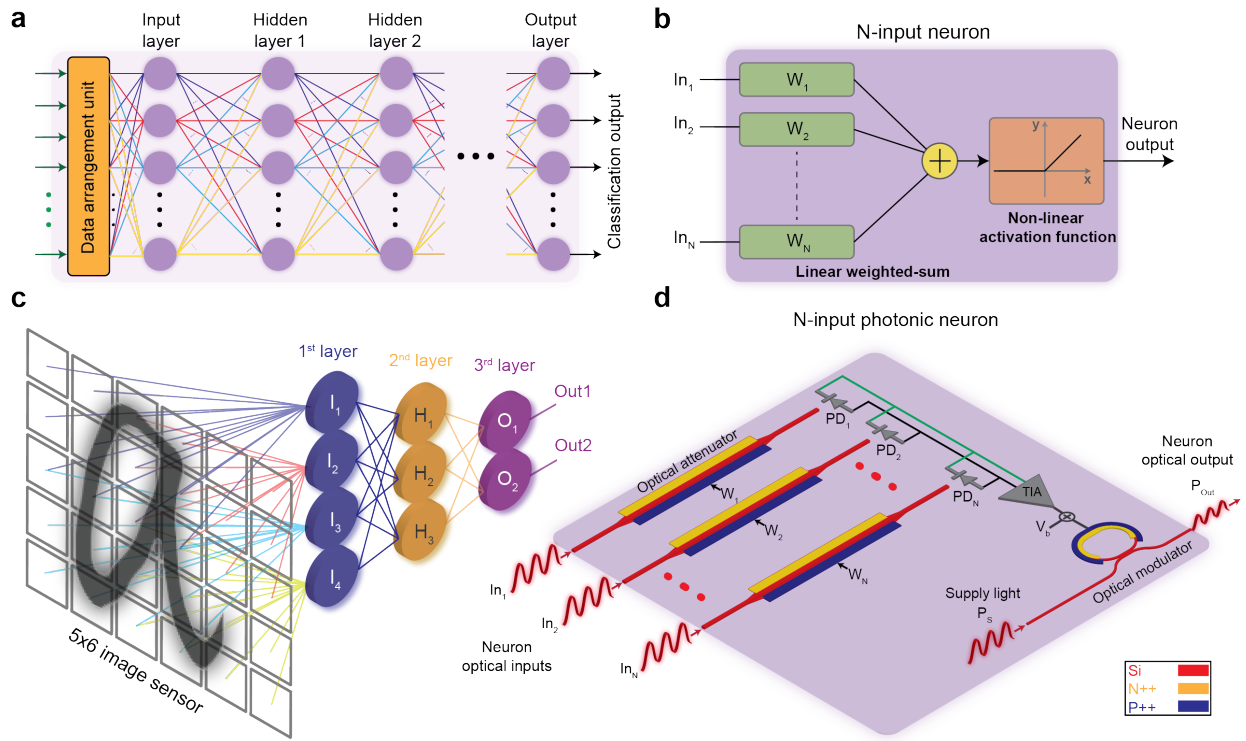

**Fig. 1 | Conventional and photonic-electronic deep neural networks. a,** Block diagram of a conventional deep neural network consisting of a data arrangement unit followed by the input layer, multiple hidden layers, and an output layer providing classification outputs. **b,** The structure of a conventional N-input neuron used in the network in **a**, where the linear weighted-sum of the inputs is passed through a nonlinear activation function to generate the neuron output. **c,** The architecture of the implemented PDNN chip, where the input image is formed on a 5x6-pixel array and is arranged into four overlapping sub-images. Each pixel of each sub-image is routed to one of the neurons of the 1st layer. The 2nd and 3rd layers are fully-connected to their previous layers. The network generates two outputs. **d,** The structure of an implemented N-input photonic neuron, where the weights of N optical input signals are adjusted using optical PIN attenuators and summed after photo-detection using parallel photodiodes. The photocurrent $i_{sum}$ is converted to a voltage and amplified using a trans-impedance amplifier (TIA). The TIA output is then used to drive an optical micro-ring modulator realizing the rectified linear unit (ReLU) nonlinear activation function, where the neuron optical output is generated by modulating the supply light.

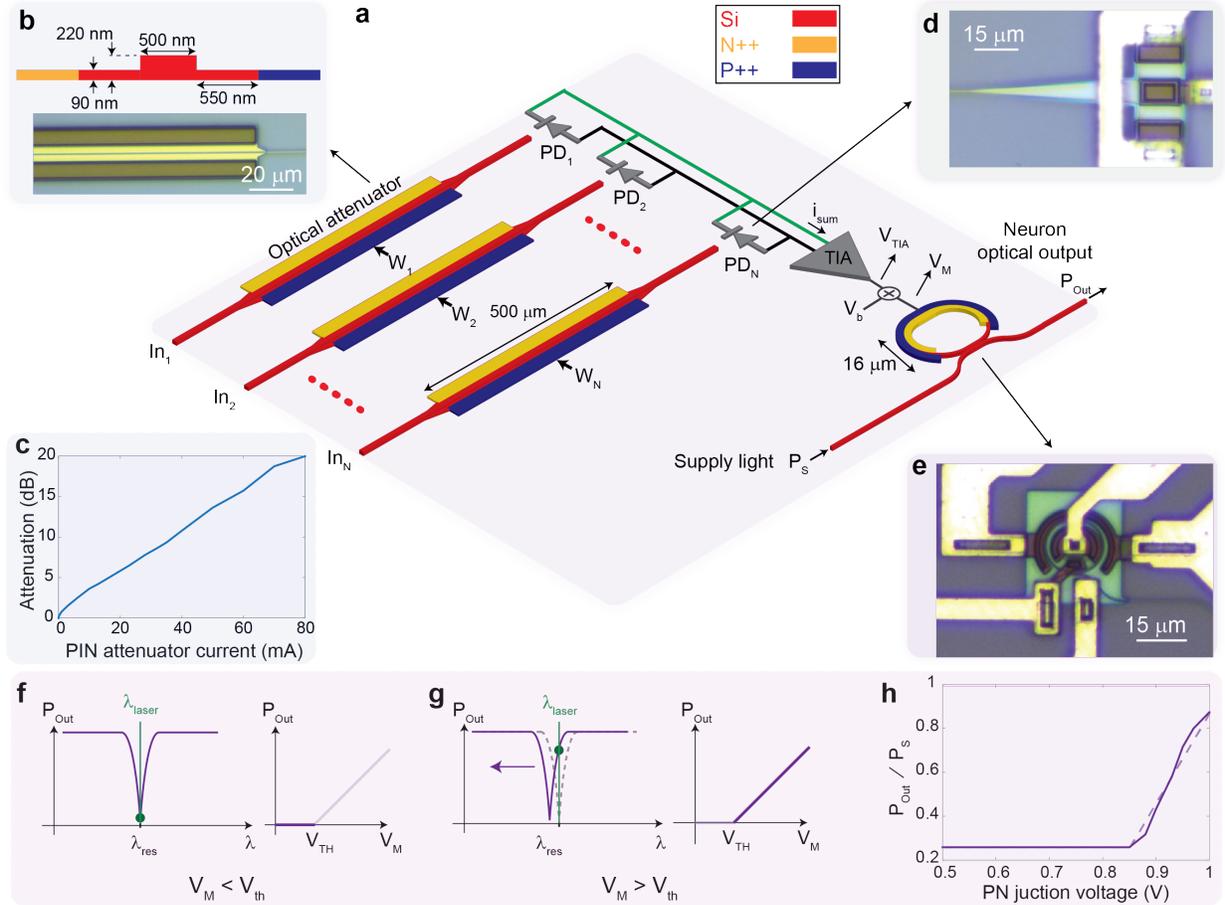

**Fig 2 | Photonic-electronic neuron implementation. a,** The schematic of the implemented on-chip photonic-electronic neuron with N optical inputs and one optical output, realized using different electro-optical devices. **b,** The cross-section and microphotograph of the P-doped-intrinsic-N-doped (PIN) attenuator, which is realized by creating P++ and N++ regions on the two sides of a nanophotonic waveguide. **c,** The attenuation of the PIN attenuator as a function of the injected current. **d,** The microphotograph of a SiGe photodiode (PD) used after each PIN attenuator. **e,** The microphotograph of the micro-ring modulator used to realize the ReLU activation function. **f,** For the case that the micro-ring is aligned with the wavelength of the supply light, while the voltage across the PN junction ($V_M$) is smaller than the turn-on voltage of the PN junction ($V_{TH}$), the junction remains off. In this case, no carrier is injected into the junction and the micro-ring resonance remains unchanged, resulting in a low neuron output power. **g,** For the case that $V_M > V_{TH}$, the PN junction turns on injecting carriers into the junction. As a result, the waveguide refractive index changes, shifting the micro-ring resonance. In this case, the neuron output power increases as $V_M$ (corresponding to the weighted-sum of the neuron inputs) increases. **h,** The measured output power of the micro-ring modulator (normalized to the supply light power) as a function of the voltage across the micro-ring PN junction, $V_M$.

The structure of a photonic neuron with N inputs ($In_i$) and one output is shown in Fig. 2a, where linear computation (*i.e.* weighted-sum of the input signals) is performed optically and the nonlinear activation function is realized opto-electronically. To calculate the weighted-sum of the neuron input signals, first, an array of 500-μm-long P-doped-intrinsic-N-doped (PIN) current controlled attenuators are used to individually adjust the optical power within each input nanophotonic waveguide of the neuron. The cross-section of the PIN attenuator as well as its microphotograph are shown in Fig. 2b, where P-doped and N-doped sections are placed on two sides of a silicon ridge waveguide. By forward biasing the PIN junction and injecting carriers, the power of the optical wave (*i.e.* the signal weight) within each neuron input can be adjusted. The optical attenuation as a function of the injected current is shown in Fig. 2c. To add weight adjusted signals within each neuron, the output of attenuators are photo-detected using silicon-germanium (SiGe) photodiodes (PDs) and the resulting photocurrents are combined to generate the weighted-sum of the neuron inputs, $i_{sum}$. The microphotograph of the SiGe PD is shown in Fig. 2d.

To generate the neuron output, the weighted-sum of the neuron inputs is passed through a nonlinear activation function. Here, the rectified linear unit (ReLU) function, offering fast convergence[11,12], is used as the nonlinear activation function and is realized by utilizing the electro-optic nonlinear response of a PN junction micro-ring modulator (Fig. 2e)[32]. In Fig. 2a, the electrical current $i_{sum}$, representing the weighted-sum of the neuron inputs, is amplified and converted to a voltage using a linear trans-impedance amplifier (TIA). The input voltage of the micro-ring modulator (driving the forward-biased PN junction), $V_M$, is generated by adding a dc voltage, $V_b$, to the TIA output voltage, $V_{TIA}$. The output of a laser coupled into the chip is uniformly distributed among all neurons within the network providing the same per-neuron supply light. The supply light routed to each neuron is coupled to the optical input of the micro-ring modulator. Consider the case that the resonance wavelength of the micro-ring modulator, $\lambda_{res}$, is initially aligned with the wavelength of the supply light, $\lambda_{laser}$. When the input voltage to the micro-ring modulator, $V_M$, is smaller than the threshold voltage, $V_{TH}$ (corresponding to the turn-on voltage of the micro-ring PN junction), the PN junction remains off and no carrier is injected into the PN junction (Fig. 2f). As a result, the resonance wavelength of the micro-ring remains aligned with the wavelength of the supply light and the output power of the micro-ring modulator (*i.e.* the neuron optical output power, $P_{out}$) remains low as the supply light is filtered by the notch response of the micro-ring modulator. When the weighted-sum of the neuron inputs, $i_{sum}$, is large enough such that $V_M$ exceeds $V_{TH}$, the PN junction turns on, changing the refractive index of the optical waveguide within the PN-junction as carriers are injected into the junction. As a result, the resonance wavelength of the micro-ring modulator shifts, increasing the neuron optical output power as shown in Fig 2g. The measured response of the micro-ring modulator configured as an electro-optic ReLU is shown in Fig. 2h, where $P_{Out}/P_S$ closely follows a rectified linear characteristic as a function of $V_M$. Note that the ReLU threshold ($V_{TH}$) can be adjusted by setting $V_b$.

**Photonic CNN image classifier chip**

Figure 3 shows the top level architecture of the implemented photonic classifier as well as the microphotographs of the main blocks. The image of the target object is formed on the 5x6 array of input grating couplers serving as the input pixel array. The coupled optical waves into the input pixels are routed to the neurons of the 1st layer of the network. The 30 signals received by the pixels of the input pixel array are split into four sets of overlapping 12-pixel sub-images, each routed to a single neuron of the 1st layer ($I_1$ to $I_4$) using a photonic distribution network designed using nanophotonic waveguides, Y-junction splitters, and waveguide crossings (Fig. 3b). To ensure uniform optical power distribution, the number of Y-junctions and crossings are balanced for all 48 optical paths that route the input pixel array to the 1st layer of the PDNN chip.

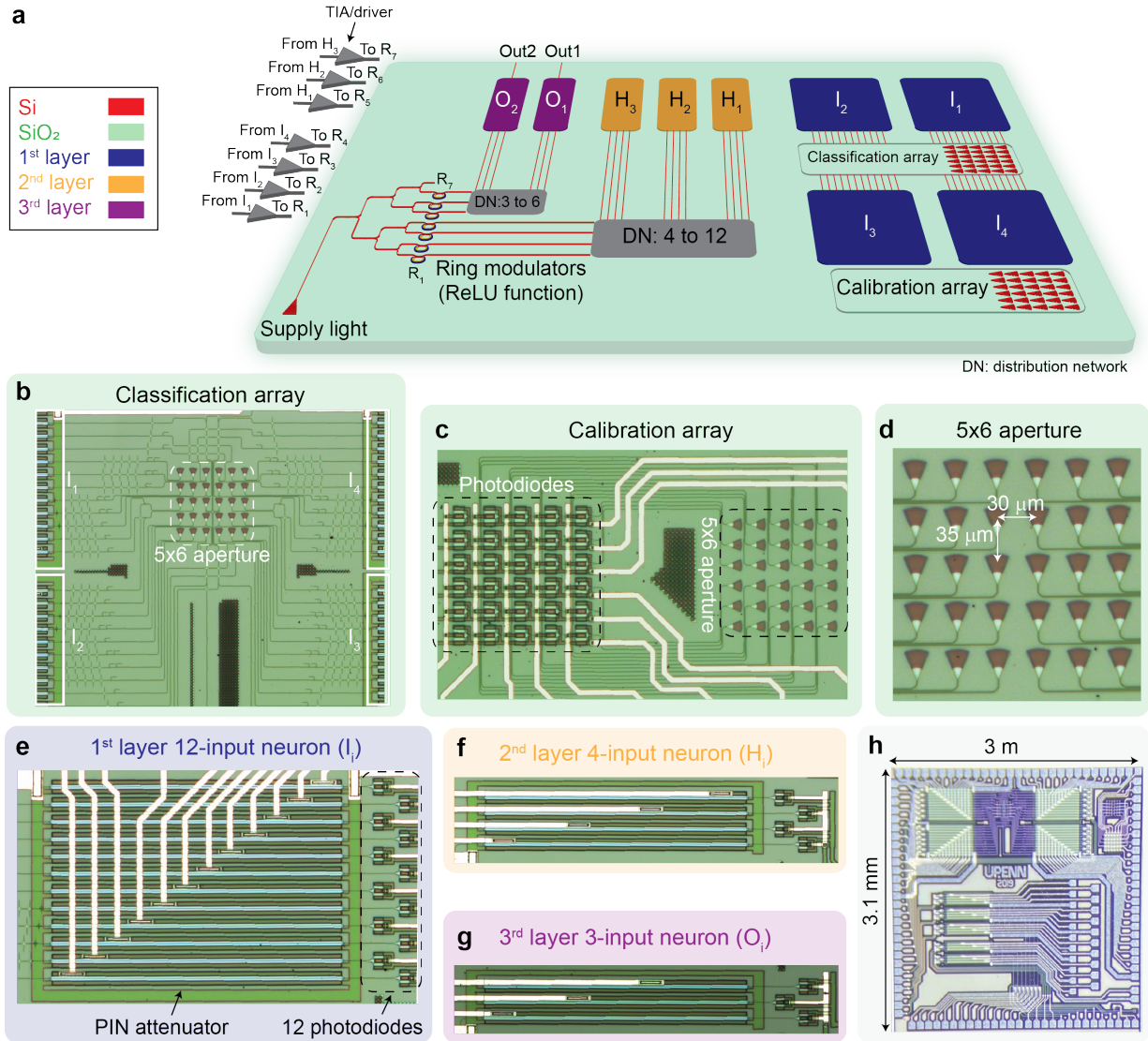

**Fig. 3 | The implemented photonic classifier chip. a,** The top-level block diagram of the PDNN chip. Two 5x6 arrays of grating couplers are used as **b,** the input pixel array and **c,** the calibration array. **d,** The 5x6 array of grating couplers showing the corresponding element pitch. The input pixel array (used for classification) generates 4 sets of 12 optical signals that are routed to the neurons of the 1st layer. The supply light is uniformly distributed among the neurons of the 2nd and 3rd layers and passes through 7 micro-ring modulators to realize the ReLU nonlinear activation function. Seven off-chip TIAs are used to drive the on-chip modulators. The system generates two outputs that are used for up to 4-class classification. **e, f, g** The microphotographs of an individual neuron within 1st, 2nd, and 3rd layers showing the PIN attenuators and the parallel PDs placed after the attenuators, respectively. **h,** The microphotograph of the photonic chip implemented in the AMF 180 nm SOI process.

A secondary identical 5x6 grating coupler array is also fabricated on the PDNN chip and used for the training of the PDNN chip and the image formation calibration, where the optical power received by each pixel is monitored using a photodetector (Fig. 3c). The microphotograph of the 5x6 pixel array is shown in Fig. 3d with an aperture size of about 140 μm by 150 μm.

After the arrangement of the input pixels to overlapping sub-images used to perform convolution, the light is processed using 3 layers; the 1st layer (input layer), which consists of four 12-input neurons, is fully connected to three 4-input neurons of the second (hidden) layer. The hidden layer is fully connected to the output layer. The 3$^{rd}$ layer (output layer) consists of two 3-input neurons. Single-mode nanophotonic waveguides with a loss of less than 0.2 dB/mm are used to connect neurons of different layers within the PDNN chip. The structures of the 1$^{st}$ ($I_i$), 2$^{nd}$ ($H_i$), and 3$^{rd}$ ($O_i$) layer neurons are shown in Figs. 3e, 3f and 3g, respectively. Outputs of $I_1$ to $I_4$ and $H_1$ to $H_3$ are connected to micro-ring modulators $R_1$ to $R_4$ and $R_5$ to $R_7$, through TIAs followed by drivers, respectively. The output layer consists of two neurons and therefore, the classifier allows for two simultaneous outputs (Out1 and Out2) that can be used for up to 4-class classification. A laser is coupled into the PDNN chip to provide the supply light to individual neuron within all layers. Figure 3h shows the microphotograph of the photonic classifier chip implemented in the AMF 180 nm silicon-on-insulator (SOI) process.

**Image classification demonstration**

The implemented PDNN chip was used to demonstrate 2- and 4-class classifications. First, in the training phase, the network was trained using a set of training images to determine the weight vectors for neurons of all layers. Then in the classification phase, a different set of testing images was classified using the trained PDNN chip. Note that while the implemented PDNN chip is designed to perform instantaneous end-to-end image classification, the network is trained using the on-chip secondary pixel array, where the training images are formed, the pixel values are recorded and a simulation platform based on Keras[31], an open-source neural network library written in Python, is used to find the weight vectors off-chip. The details of the training process are presented in the Methods section (and Extended Data Fig. 3).

After the PDNN is trained, the resulting weight vectors are transferred back to the PDNN chip to perform real-time classification of target objects in the test set and quantify the classification accuracy. Figure 4a shows the schematic of the measurement setup used to demonstrate image classification. Two laser sources are used; laser 1 is used for image formation on the classification/calibration arrays, and laser 2 is used to provide supply light for neurons. The output power of laser 1, emitting at 1532 nm, is amplified using an erbium-doped fibre amplifier to approximately 63 mW and coupled to an optical collimator with a beam diameter of 870 $\mu$m. The collimated beam illuminates the object plane that consists of printed letters on a transparency film. The printed letters are attached to a custom fabricated Plexiglas holding frame which is mounted on a high precision XY positioning system with a resolution of better than 1 $\mu$m. The collimated beam passes through the object plane forming the image of the target object on the 5x6 classification array.

Laser 2 emits 2.5 mW at 1559.93 nm. Control loops are utilized to achieve and maintain correct alignment of the micro-ring modulators in presence of thermal and fabrication process variations resulting in the reliable realization of a rectified linear function. Once all the weights are set and the chip reaches thermal equilibrium, the alignment control loop is engaged to thermally tune the ring modulators such that all resonance wavelengths are aligned with the wavelength of laser 2. The details of the alignment algorithm are presented in the Methods section.

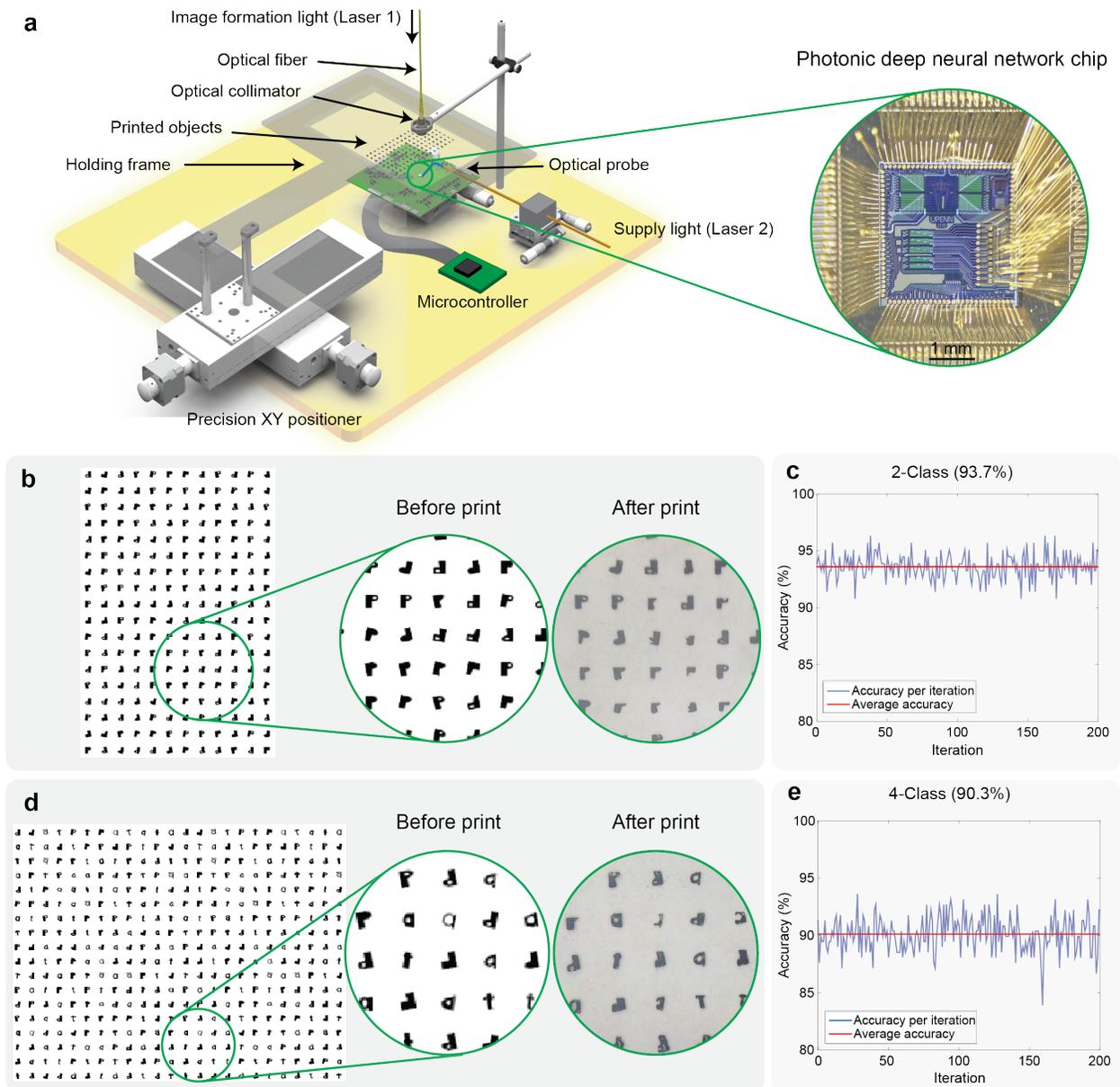

**Fig. 4 | Image classification demonstration. a,** Classification measurement setup. Laser 1 provides the light for image formation on the input pixel array (in the classification phase) or the calibration array (in the training phase) while laser 2 is used as the supply light. The target objects (dataset) are printed on a transparency film which is mounted on a custom fabricated frame. A high precision XY positioner is used for scanning through the dataset. A microcontroller is used to write the weights into the photonic chip and to implement micro-ring modulator alignment control loops. **b,** The dataset consisting of letters "p" and "d" for the 2-class classification measurements. The zoomed-in views show the letters before and after printing. **c,** Measured accuracy as a function of the number of iterations. In each iteration, a random set of measured data is used to find the thresholds to separate the classes showing the robustness of the classification algorithm (see Extended Data Fig. 3 for details of training and threshold calculations). **d,** The dataset consisting of letters "p", "d", "a", and "t" for 4-class classification measurements and the zoomed-in views of the before and after print letters. **e,** Classification accuracy results for the 4-class measurements.

After all micro-ring modulators are aligned, the system sequentially goes through the test images (all printed on the same transparency film) using the XY positioner, while the voltages of Out1 and Out2 are continuously monitored. As each image is formed on the input pixel array, the result appears at the PDNN chip output almost instantaneously. The details of the control circuit are presented in the Methods section.

After instantaneous classification, the optical output of the output layer, representing the estimated class of the target object, is photo-detected on chip. The photocurrents are then wire-bonded off-chip and converted to a voltage. A subset of the measured data (output voltage) is used to find optimal threshold values to separate different classes. These threshold values are then used on all images in the test set to label the output classes using a simple linear process. The classification accuracy was calculated by comparing the detected classes to the reference data labels. In order to demonstrate the robustness of this algorithm, this process is repeated for 200 iterations on randomly selected subset of samples using the cross-validation method[33]. Then, the average classification accuracy over 200 iterations is reported. The details of the classification accuracy calculations and the threshold determination is presented in the Methods section.

Two measurements were conducted to demonstrate the functionality of the PDNN chip. First, in a 2-class case, a dataset consisting of 216 letters, 108 "p" and 108 "d", was generated and printed on a transparency film using a commercial laser printer. Figure 4b shows a picture of the dataset, where different forms of each letter are designed to increase the variety of the dataset. Note that the limited print resolution adds an extra level of random variations to the dataset, making the classification more challenging. This is shown in the zoomed-in view in Fig. 4b. Figure 4c shows the measured 2-class classification accuracy as a function of the number of iterations, where an average classification accuracy of 93.7% is achieved.

In the second experiment, a dataset consisting of 432 letters of "p", "d", "a", and "t", 108 of each, was generated to demonstrate 4-class classification. Figure 4d illustrates the designed dataset and the corresponding printed version. The accuracy for 200 iterations is plotted in Fig. 4e, where an average accuracy of 90.3% is achieved. These results show that even with larger number of classes (*i.e.* letters) and in presence of printer induced variations and noise, the PDNN chip still achieves a high classification accuracy. As a point of comparison in terms of classification accuracy, we used a standard conventional neural network (CNN), implemented in Python environment using Keras[31], to classify the same printed 4-class dataset. The standard CNN architecture has been previously used for classification of MNIST[34] handwritten digits dataset to achieve accuracies of higher than 99%[35] and is tailored to our 4-class dataset. This significantly larger network (with more than 190 neurons) compared to the reported PDNN chip, achieves a classification accuracy of about 96% for the printed 4-class ("p", "d", "a", and "t") dataset used in this work.

**Discussion**

In general, the classification speed of the proposed PDNN chip is mainly limited by the bandwidths of the micro-ring modulators, the SiGe photodiode, and the TIA, since the processing is performed as the waves propagate within the chip. Using commercial SOI fabrication processes, the overall bandwidth of tens of gigahertz can be achieved for these blocks. Since the propagation delays for these blocks are inversely proportional to their bandwidths, sub-100 picosecond processing speed can be achieved for each neuron layer. The ring modulator and the SiGe photodiodes that are used on the implemented PDDN chip both have 3-dB bandwidth of larger than 30 GHz[18]. As a result, the implemented PDNN chip is capable of performing end-to-end classification within a single clock cycle of a state-of-the-art GPU with a clock frequency of 3 GHz[36].

The implemented PDNN chip can be scaled to a classifier with a larger number of pixels in order to instantaneously classify higher resolution images and much more complex patterns. The complexity of routing overlapping sub-images to the neurons of the input layer (to perform convolution), which is an important scaling challenge, can be addressed either by using a fabrication process with multiple photonic routing layers[37] allowing for more complex photonic routing, and/or through tiling, where multiple pixel arrays are placed next to each other.

In summary, we have demonstrated the first end-to-end photonic deep neural network classifier chip that performs instantaneous image classification through computation by propagation of optical waves, eliminating the need for an image sensor, digitization, and large memory modules. Benefiting from the large bandwidth and low propagation loss offered by integrated photonic platforms, the implemented chip enables realization of neural networks that are significantly faster and consume less power compared to the conventional all-electrical solutions. Low energy consumption and ultra-low computation time offered by our photonic classifier chip can revolutionize applications such as event-driven and salient object detection[38,39] both as a standalone classifier or in conjunction with electronic processors utilizing the instantaneous classification of the PDNN chip as well as the re-configurability and flexibility of electronic processors.

## Methods

**Image formation uniformity and path loss**

Using the calibration array we can verify the image formation quality. One important consideration is the uniformity of the image. To check that, we uniformly illuminate the chip, with no obstruction, while measuring the photocurrent of individual pixels in the calibration array. In this case, the measured non-uniformity is less than 5%, which is low enough for the classification system to extract the features properly.

The same measurement can be used to estimate the path loss from the optical collimator to the power coupled into the chip by each pixel (grating coupler). In the case that the power coupled to the collimator is about 63 mW, the measured photocurrent of each photodiode is about 3 μA. The responsivity of the photodiode is about 0.8 A/W. Therefore the power coupled into the waveguide connected to each grating coupler is estimated to be 4 μW. This result in a total path loss of about 42 dB. This loss is mainly due to 3 factors; (1) the overlap between the input pixel array aperture area and the beam spot, which can be written as

$$\frac{A_{aperture}}{A_{beam}} = 0.035 \, (\approx 14.5 dB),$$

(2) aperture fill factor (*i.e.* the area of each grating coupler relative to the aperture area)

$$Fill \ factor = \frac{Pixel \ area}{Aperture \ area} = 0.0048 \, (\approx 23 dB),$$

and (3) the grating coupler measured loss of about 5 dB. Note that the transmission coefficient of the transparency film is almost one.

**Micro-ring modulator alignment algorithm**

In the implemented PDNN chip, 7 micro-ring modulators are used to implement and approximate the neural ReLU nonlinear activation function; four micro-ring modulators at the output of the 1$^{st}$ layer and three micro-ring modulators at the output of the 2$^{nd}$ layer. As discussed earlier, a supply light is coupled into the optical input of each micro-ring modulator. Since the wavelength of the supply light is the same for all micro-rings, the resonance wavelengths of all micro-ring modulators must be aligned to ensure reliable and repeatable realization of ReLU functions for all neurons. In practice, the resonance wavelength of micro-ring resonators may vary due to the fabrication process variations and temperature change. Therefore, in addition to a careful design and layout of the micro-rings, control loops were implemented to compensate for any misalignments between the resonance wavelength of micro-ring modulators and the wavelength of the supply light. Each micro-ring modulator can be tuned using an N-doped heater section (serving as a thermal phase shifter) with a measured resistance of about 1.9 kΩ. Extended Data Fig. 1a illustrates the algorithm used to perform the micro-ring alignment. First, the supply light is switched on and all weights are set; properly setting the weights is of particular importance as biasing the PIN attenuators (to adjust the weights within neurons) may increase the temperature of the chip. Therefore, micro-ring alignment process should be performed while the weights are set and the chip has reached thermal equilibrium. In this case, the control loop sequentially adjusts the heater voltages to minimize the difference between the sum of the outputs of the neurons of the 2$^{nd}$ and 3$^{rd}$ layers (*i.e.* H$_1$ to H$_3$, O$_1$, and O$_2$), V$_{SUM}$, and a reference voltage, V$_{REF}$. In addition, the algorithm ensures that the heater voltages do not exceed the maximum allowable value, V$_{max}$. At the end of each iteration (*i.e.* after the adjusting the voltage of all heaters), if V$_{SUM}$ becomes smaller than V$_{REF}$, then V$_{REF}$ is set to V$_{SUM}$ and the next iteration starts. Once all rings are aligned, the optimal heater biasing voltages are recorded and used during the classification process.

To verify the performance of the ring alignment control loop, first the laser that illuminates the 5x6 input pixel array is turned off (Extended Data Fig. 1b). In this case, the outputs of the neurons of the 1$^{st}$ layer (I$_i$) are zero. Since micro-rings are properly aligned, the outputs of the neurons of the 2$^{nd}$ and 3$^{rd}$ layers (the corresponding the ReLU function output) remain low. Then, the input laser is turned on, uniformly illuminating the 5x6 input pixel array. In this case, I$_1$ to I$_4$ increase, shifting the resonance wavelengths of the micro-ring modulators, which results in a large change in the outputs of the neurons of the 2$^{nd}$ layer, H$_1$ to H$_3$. Similarly, the output of the neurons of the 3$^{rd}$ layer, O$_1$ and O$_2$, will change. The output voltages of neurons of different layers before and after uniform illumination of the input pixel array are shown in Extended Data Figs. 1b and 1c.

**Electronic control circuitry**

Extended Data Fig. 2 shows the block diagram of the electronic system used to control and drive the photonic components of the classifier chip. The circuit consists of a microcontroller utilized to generate the data and clock signals for the serial digital-to-analogue converter (DAC) array to set the weights, thermally align the micro-ring modulators and adjust the threshold voltage of each ReLU block. A serial interface is used to write the data into the serial DAC array. There are 66 PIN attenuators on chip to set the weights corresponding to the neurons (4x12 in the 1$^{st}$ layer, 3x4 in the 2$^{nd}$ layer, and 2x3 in the 3$^{rd}$ layer), seven heaters were used for thermal tuning of the micro-rings, and seven bias voltages (V$_b$) to adjust the threshold of the ReLU blocks.

**PDNN chip training process**

Prior to performing the image classification on the test set, the PDNN chip was trained to find the optimal weight vectors. First, the image of each letter in the training set was formed on the on-chip secondary (calibration) pixel array and the corresponding pixel values were recorded. Then, to find the optimal weight vectors, the recorded pixel values for each image of the training set were fed into a digital neural network implemented in Python using Keras[31]. The architecture of this digital neural network is identical to that of the PDNN chip with ReLU nonlinear activation function. The training and weight optimization were performed using the stochastic gradient descent algorithm[9].

While the PDNN chip with two outputs can be used in a conventional way to classify a 2-class data set, a simple additional step enables the implemented PDNN chip to perform classification of datasets with a larger number of classes. In this case, a simple linear combination of two outputs of the 3$^{rd}$ layer, V$_{out}$ = Out1 – Out2, can be formed and compared with one or a set of threshold values to determine the class of each input image. Therefore, the training process also includes the calculation of the threshold values required to optimally separate different classes. In this work, one and three threshold values were used for the 2- and 4-class cases, respectively.

During the training phase, for the algorithm to find optimal threshold values, a subset of the data is used. As shown in Extended Data Fig. 3a, as the measured data values are fed to the algorithm, the threshold values are constantly updated and move closer to the optimal values resulting in higher classification accuracies. The optimum weight vectors were translated to the corresponding input voltages of the PIN attenuator array using a look-up table (containing the amount of attenuation as a function of the attenuator input voltage). A microcontroller followed by an array of digital-to-analogue converters were used to write the optimum weight vectors into the PDNN chip. During the classification phase, the threshold values calculated during the training process were used.

Extended Data Figs. 3b and 3c show the classification accuracy as a function of the number of measured data used to determine the threshold values in the training phase. As shown, the accuracy increases and converges to its maximum value as more data is fed into the algorithm and more accurate threshold values are calculated. The calculated threshold values depend on the sequence of the input data. Therefore, to ensure the robustness of the threshold calculation algorithm, this process is repeated multiple times. The error bars in Extended Data Figs. 3b and 3c correspond to the variation in accuracy as a function of the number of input data points. Based on these graphs, 25% of the 2-class dataset and 50% of the 4-class dataset were used to calculate the corresponding threshold values. The remaining data in both cases was used in the classification phase and the resulting accuracies are shown in Figs. 4c and 4d, respectively.

**Chip fabrication**

The photonic chip was fabricated in the AMF 180 nm SOI process with a 2 $\mu$m thick buried oxide. Single-mode 220 nm thick and 500 nm wide nanophotonic waveguides with a loss of less than 2 dB/cm were used for photonic routing. The grating couplers used in the input pixel array as well as in the calibration array have a measured coupling efficiency of about 30% while the grating coupler used for coupling the supply light has a measured coupling efficiency of about 40%. The measured excess loss of the Y-junctions and the loss of waveguide crossings used in the distribution network (following the input pixel array) are about 0.5 dB and less than 0.1 dB, respectively. The photodiodes have a measured responsivity of about 0.8 A/W and a 3-dB bandwidth greater than 30 GHz. The ring modulators have a 3-dB bandwidth of more than 30 GHz.

**Data availability**

The data that support the plots and other findings within this paper are available from the corresponding author upon reasonable request.

**Acknowledgements**

This work was supported by Office of Naval Research of the United States under the award number N00014-19-1-2248.

**Author contributions**

F. Ashtiani and F. Aflatouni conceived the design idea. F. Ashtiani designed, simulated, and laid out the photonic chip. F. Ashtiani and A.J.Geers conducted the measurements. F. Aflatouni supervised the project. F. Ashtiani and F. Aflatouni wrote the manuscript.

**Competing interests**

The authors declare no conflicts of interest related to this article.

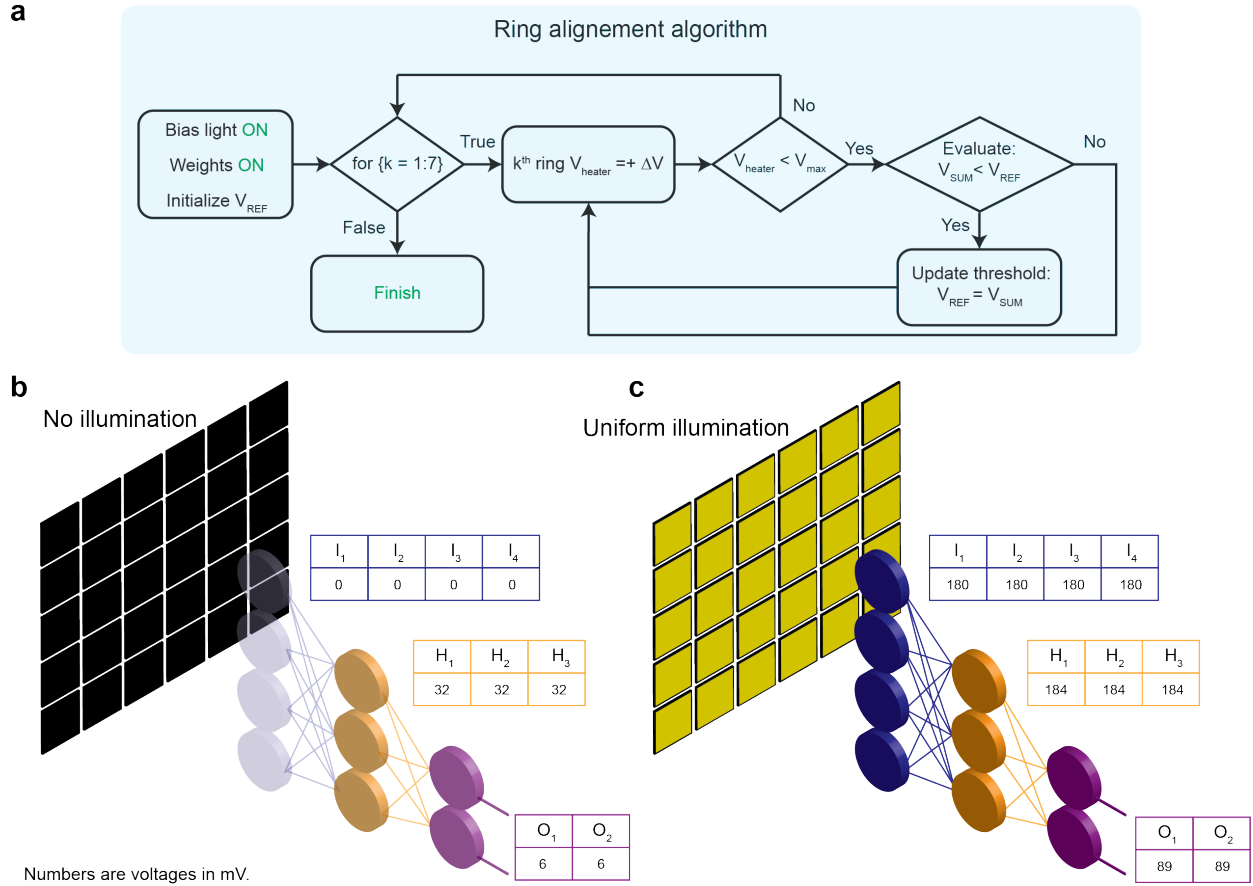

**Extended Data Fig. 1 | Micro-ring alignment algorithm and characterization. a,** The implemented algorithm flowchart for micro-ring alignment. The cost function to be minimized is $V_{SUM}$, which is the sum of the outputs of the 2$^{nd}$ and 3$^{rd}$ layers (*i.e.* $H_i$ and $O_i$). All micro-rings are thermally tuned in order to find the optimal heater voltages that correspond to the same resonance wavelengths for all 7 rings. **b,** In case of no input illumination, the output of the neurons of the 1$^{st}$ layer ($I_i$) are zero. If micro-rings are properly aligned, the outputs of the neurons of the 2$^{nd}$ and 3$^{rd}$ layers remain low. **c,** In the case that the optical input is uniformly illuminating the input pixel array, if all rings are aligned, $I_1$ to $I_4$ will increase, shifting the resonance wavelengths of the micro-ring modulators, which results in a large change in the outputs of the neurons of the 2$^{nd}$ and 3$^{rd}$ layers.

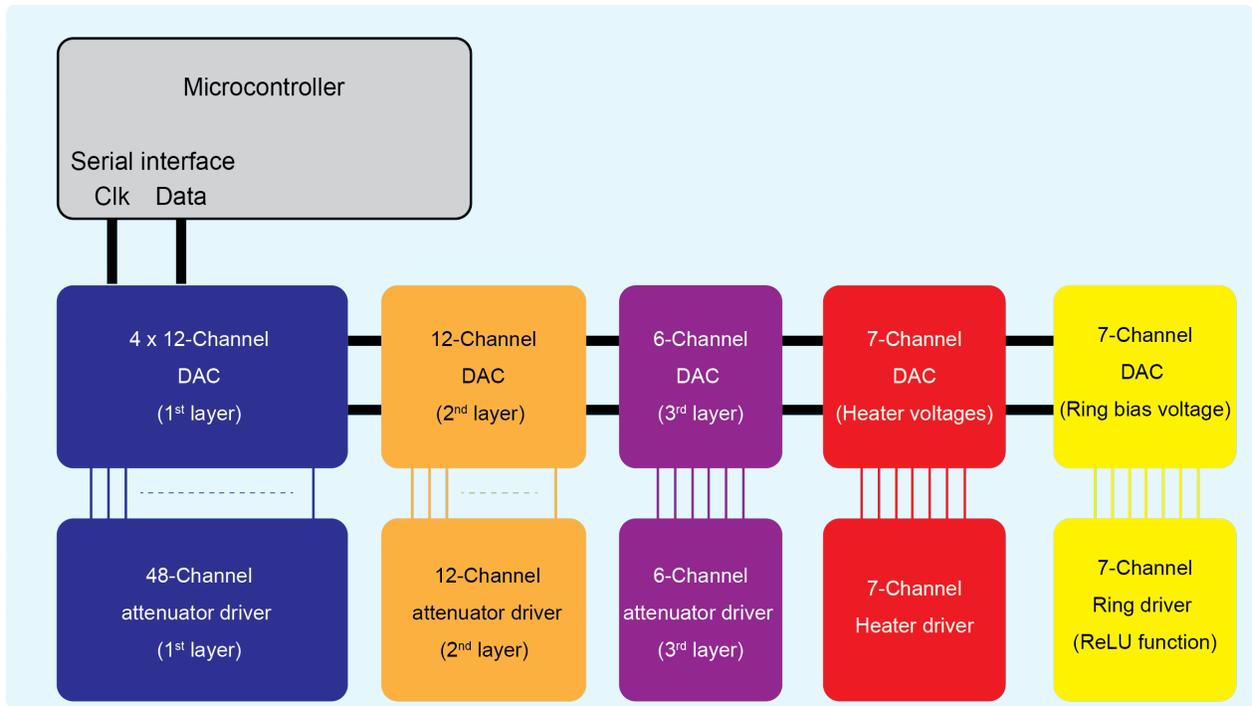

**Extended Data Fig. 2 | Electronic control circuit block diagram. a,** The microcontroller sends the clock and data signals to the serial digital-to-analogue converters (DACs) while the outputs of the DACs are connected to their corresponding drivers to drive the on-chip photonic devices (PIN attenuators, ring PN junctions and micro-ring thermal phase shifters).

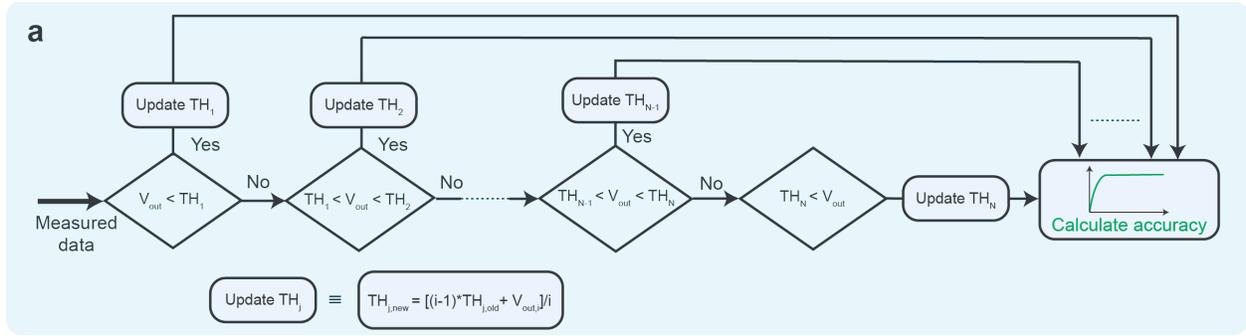

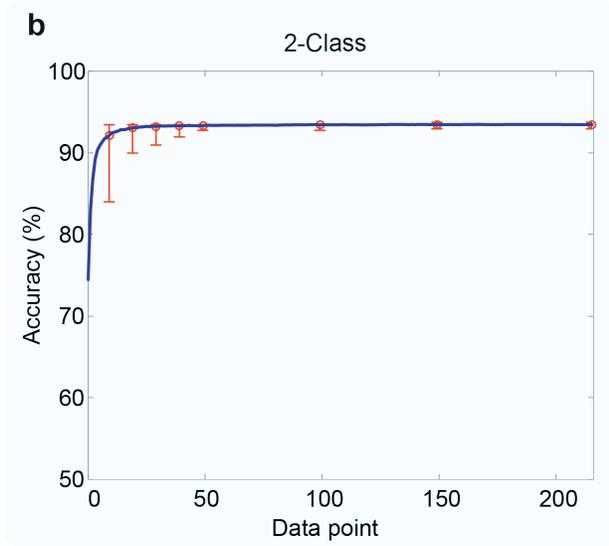
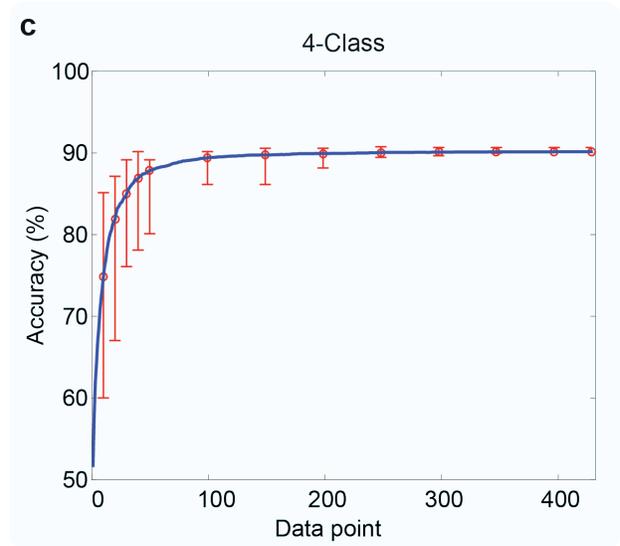

**Extended Data Fig. 3 | PDNN chip training and threshold calculations. a,** The implemented algorithm to find and update the threshold values to properly separate N different classes. A linear combination of the network output, in this case, the differential output defined as $V_{out}$ = Out1 – Out2, is measured and compared with different threshold levels. The threshold values ($TH_j$) are updated one by one as measured network differential output values ($V_{out,i}$) are sequentially passed into the algorithm. The classification accuracy is calculated and plotted as the threshold values are updated. The classification accuracy as a function of input measured data stream is plotted in **b,** for the 2-class and **c,** for the 4-class cases, respectively. The error bars show the variations in the classification accuracy for specific number of data points used for calculating the threshold values. In this case, the horizontal axis in **b** and **c** represents the number of data points that were randomly selected from all data points and used to determine the threshold. This threshold is then used to classify all data points and the classification accuracy is calculated. This process was repeated 20 times for each selected number of data points and the resulting range of accuracies are shown with error bars.

**Extended Data Table 1| List of equipment and devices**

| Equipment | Model |
|---|---|
| Laser 1 | HP 8168F |
| Laser 2 | Agilent 81682B |
| XY positioner and controller | Thorlabs NRT150 and BSC103 |
| Driver op-amp | Texas Instruments TLV3544 |
| Digital to analogue converter (DAC) | Analog Devices AD8802 |
| Microcontroller | ATMEL ATSAM3X8E |
| Optical collimator | Thorlabs CFC-5C |
| Polarization controller | Thorlabs FPC-31 |
| Erbium-doped fibre amplifier (EDFA) | Optilab EDFA-I24-B |
| DC power supply | Agilent E3646A |